\documentstyle[12pt,psfig]{article}
\begin{document}
\title{ Tunneling of the Closed Friedmann Universe
 with Generation of Scalar  Waves}

\author{ V. Ts.  Gurovich\thanks{Physics Institute of NAN KR,
265 a, Chui str., Bishkek, 720071,  Kyrgyz Republic, e-mail
gurovich@grav.freenet.bishkek.su}, H.-J. Schmidt\thanks{Potsdam Univ.,
 Inst. f. Math.,
Germany and Inst. f. Theoretische Physik, Freie Univ. Berlin,
Germany, hjschmi@rz.uni-potsdam.de} and I. V. Tokareva
\thanks{KRSU, 44, Kievskaya st., Bishkek,
720000, Kyrgyz Republic, e-mail astra@freenet.kg}}

\maketitle

\begin{abstract}

The evolution of the closed Friedmann Universe with a packet of
short scalar waves is considered with the help of the
Wheeler--DeWitt equation. The packet ensures conservation of
homogeneity and isotropy of the metric on average. It is shown
that during tunneling the amplitudes of short waves of a scalar
field can increase catastrophically promptly if their influence to
the metric do not take into account. This effect  is similar to
the Rubakov-effect of catastrophic  particle creation calculated
already in 1984.

In our approach
 to  the problem it is possible to consider self-consistent dynamics of the
expansion of the Universe and amplification of short waves.
 It results in a  decrease of the barrier and interruption of amplification of waves, and we get an exit of the
 wave function  from the quantum  to the  classically available
region.
\end{abstract}

\par

\section{ Statement  of the   problem}

Here we present the basic equations of the problem.

In the space-time of the closed cosmological Friedmann models
with space-time metric
\begin{equation}
\label{doc1} ds^2=dt^2- a^2(t)[d\chi ^2+\sin ^2\chi (d\theta
^2+\sin ^2\theta d\varphi ^2)],
\end{equation}
where $0\le\chi, \theta \le \pi, \quad 0\le\varphi \le 2\pi $ ,
the action is assumed to be
\begin{equation}
\label{doc2} \qquad \quad S=\int\left[-\frac{R}{16\pi G}
-\stackrel{-}{\Lambda } +\frac{1}{2}\varphi^{,i}\varphi _{,i} -
\frac{1}{2}m^2\varphi^2\right]\sqrt{-g}\,d\Omega,
\end{equation}
Here $R$ is a scalar curvature, $\varphi $ is a potential of the
scalar quantum field with a mass $m$, ($\hbar=c=1$),
$\stackrel{-}{\Lambda}$ is a cosmological $\Lambda$ - term.

The equation for $\varphi$ has the  form $\Box \varphi+
m^2\varphi=0$. Below let us set the scalar field as radial standing
waves with time-dependent amplitudes
\begin{equation}
\label{doc3} \varphi_k={\rm A}_k(t) f_k(\chi); \qquad
f_k(\chi)=\frac{\sin k\chi}{\sin \chi }
\end{equation}
Note that the functions $ f_k(\chi)$ do not have singularities on the poles $(\chi=0,
\pi)$ for integer  numbers $k$, and that  they are orthonormal
 on the coordinate $\chi$ in the measure defined by metric  (\ref{doc1}),
i.e.,
\begin{equation}
\label{doc4}
\int \sin^2 \chi f_k f_n\,d\chi=\delta_{kn} \ .
\end{equation}
\par
Now  the action
(\ref{doc2})  can be integrated  over the  angular variables,  and we
 rewrite it  after replacement of $R$ by  its value through $a$ as
\begin{equation}
\label{doc5}
 S=\frac{3\pi}{4G}\int\left[\sum_{k} \left( a^3 \stackrel{.}{\rm A}_k^2-k^2 a {\rm A}_k^2 - m^2a^3{\rm A}_k ^2 \right)-\Lambda a^3 -\stackrel{.}{a}^2 a+a \right]dt,
\end{equation}
where $$ {\rm A}_k=\sqrt{8G/3}{\rm A}_k(t); \qquad \Lambda=8\pi G
\stackrel{-}{\Lambda}/3. $$
\par
These redefinitions are made to get $G$ and $\pi$ only in front of the integral, and the
integrand to be free of both constants.

In this formula, the point denotes the  derivative with respect to $t$,  and one
supposes that all $k\gg1$, i.e., we assume that waves with wavelength comparable to the
diameter of the Universe are negligible for the dynamics.

The waves packet $(\sum_{k})$ ensures the
homogeneity of the model during expansion with a high  accuracy.
After introduction of generalized momenta the Hamiltonian function
can be presented  as
\begin{equation}
\label{doc6}
 H=-\frac{P_a^2}{4a}-a + \Lambda a^3 +\sum_{k} \left(\frac{q_k^2}{4a^3}+ k^2{\rm A}_k^2 a+
m^2a^3{\rm A}_k ^2 \right)=0.
\end{equation}
Here $ P_a, a$ are a generalized momentum and a coordinate for the
Universe as entire; $q_k$ and $ {\rm A}_k $ are the same for each
radial mode of the field. Corresponding to eq. (\ref{doc6}) the
Wheeler-DeWitt equation for the  wave function of the Universe (WF)  has
the  form

\begin{equation}
\label{doc7}
 \frac{\partial^2\Psi}{\partial a^2}-\frac{1}{a^2}\sum_{k} \frac {\partial^2\Psi}{\partial {\rm A}_k
^2}-V\Psi=0,
\end{equation}
where the superpotential $V$ is given as
\begin{equation}
\label{doc8} V=4\left[ a^2 -\frac{p}{8a^2}\left
(1-\frac{p}{2}\right)-\sum_{k}k^2 {\rm A}_k ^2a^2-\Lambda
a^4\right] .
\end{equation}
Here we are neglecting $\sum_{k}m^2 {\rm A}_k ^2a^4$ in comparison
with  the sum residual in eq. (\ref{doc8}) since $k\gg1$.
 This is justified because in the short-wavelength approximation we are
applying here, the mass of the particles does not have much influence.

The factor
ordering parameter $p$ \cite{Linde} is also introduced in eq. (\ref{doc8}).

\section{ Semiclassical approximation }
We are going to search a solution of equation (\ref{doc7}) in
semiclassical approximation
\begin{equation}
\label{doc9} \Psi_c=\exp(iS_c),
\end{equation}
where $S_c$ is a classical action. In this approximation we have
\begin{equation}
\label{doc10} \left(\frac{\partial S_c }{\partial
a}\right)^2-\frac{1}{a^2}\sum_{k}\left( \frac {\partial S_c
}{\partial {\rm A}_k }\right)^2+V=0,
\end{equation}
\par

This nonlinear equation of the first order is similar to the  Hamilton-Jacobi
equation of  analytical mechanics. For its
solution one can take advantage of the  complete integral \cite{LL}. It
can be found from the system of characteristic equations \cite{Kamke}.
As the eq. (\ref{doc10}) supposes the solution for $(\partial S
/\partial a)$, the scale factor becomes an argument along the
characteristics. From here we have
\begin{equation}
\label{doc11} \frac{d S_c }{d a}= -\frac{V }{F}; \qquad \frac{d
{\rm A}_k }{d a}= -\frac{q_k }{a^2F};\qquad \frac{dq_k }{d a}=
\frac{4k^2{\rm A}_k a^2}{F}.
\end{equation}
Here $V$ is indicated by (\ref{doc8}), and $$ F=-\sqrt
{(\sum_{k}q_k^2)/a^2-V}. $$

Now the  evolution of the WF  $\Psi$ goes along a characteristic.
Turning points  are the roots of $V(a)=0$. They separate
classically available and forbidden regions along the
characteristic eq. (\ref{doc11}) \cite{Gurovich}. A similar approach
for the strongly anisotropic WF in a Bianchi type I model is presented
 in ref. \cite{Bachmann}, and for a closed Friedmann model in ref.
 \cite{Folomeev}.

In the region under the barrier  the semiclassical solution can be
found by an ansatz of the form
\begin{equation}
\label{doc12}
\Psi=\exp(-|S_e|).
\end{equation}

The equation for $ S_e $ can be obtained from eq. (\ref{doc10}) by
changing $V\to-V$. Thus, as distinct from eq. (\ref{doc11}) the system
of the characteristics  has the  form

\begin{equation}
\label{doc13} \frac{d S_e }{d a}=\frac{V }{F_e}; \qquad \frac{d
{\rm A}_k }{d a}= -\frac{q_k }{a^2F_e};\qquad \frac{dq_k }{d a}=
-\frac{4k^2{\rm A}_k a^2}{F_e};
\end{equation}
$$\qquad F_e=-\sqrt {(\sum_{k}q_k^2)/a^2+V}.$$

\section{ The evolution of the scalar field  modes in the classically
allowed  region}

Let us investigate how amplitudes of short wavelength  change at the above
mentioned evolution of the WF. Equations for $q_k$ and ${\rm A}_k$
follow from (\ref{doc11})
\begin{equation}
\label{doc18} \frac{d {\rm A}_k }{d a}= -\frac{q_k }{a^2F};\qquad
\frac{dq_k }{d a}= \frac{4k^2{\rm A}_k a^2}{F},
\end{equation}
where $$ F=-\sqrt {(\sum_{k}q_k^2)/a^2-V}. $$

This system is an essentially non-linear one since co-factors of $q_k$
and
${\rm A}_k$ in the
right hand sides of the equations depend on $a$, too.
 However, the presence  of the large  parameter $k^2$
allows finding a solution.

Following from the system (\ref{doc18}), the equation for
the generalized momenta $q_k$ is
\begin{equation}
\label{doc19} \frac{d^2q_k }{d a^2}=
-\frac{4k^2q_k}{F^2}+\left(\frac{2}{a}-\frac{F'}{F}\right)
\frac{dq_k }{d a},
\end{equation}
where $'=d/da$. Here  $F(a)$ is  an  unknown function according to
eq. (\ref{doc18}).

It is easy to show that the following solution
\begin{equation}
\label{doc20} q_k=\frac{C_{k0} \ a}{\sqrt{2k}}\cos\Phi(a);\qquad
\Phi(a)=2k\int\limits_0^a\frac{da}{F}
\end{equation}
satisfies  the last equation with the  high accuracy of  ${\rm
 O }(1/k^2)$. And correspondingly,
\begin{equation}
\label{doc21} {\rm A}_k=-\frac{C_{k0}}{2\sqrt{2}ak^{3/2}
}\sin\Phi(a),
\end{equation}
where $C_{k0}$ is an arbitrary constant, and the phase of the
solution is chosen from the requirement of amplitude  finiteness
at $a\rightarrow 0$. The mentioned accuracy of solutions requires
the assumption that the logarithmic derivative $F'/F$ does not give a
large parameter $k$. The last statement
 follows from the fact that  each
mode of wave enters into eq. (\ref{doc18}) for $F$ as a term
\begin{equation}
\label{doc22}
 \frac{q_k^2}{a^2}+4k^2{\rm A}_k^2a^2=\frac{C_{k0}^2}{2k}=const.
\end{equation}
It is clear from substitution eqs. (\ref{doc20}), (\ref{doc21}) in
(\ref{doc22}). With taking into account (\ref{doc22}) after
rewriting of the amplitude $C_{k0}/\sqrt{2k}=C_k$, we have for $F$
the following  expression
\begin{equation}
\label{doc22a} F=-\sqrt {\sum_{k}C_k^2-4\left[a^2-\Lambda
a^4-\frac{p}{8a^2}\left (1-\frac{p}{2}\right)\right]},
\end{equation}
Further, we suppose that $\Lambda\ll 1$ and we have in the
superpotential
\begin{equation}
\label{doc22b} V\simeq4\left[a^2-\frac{p}{8a^2}\left
(1-\frac{p}{2}\right)-\frac{1}{4}\sum_{k}C_k^2\sin^2\Phi(a)\right].
\end{equation}

Values of $V$ and $F$ define an action of  WF $\Psi(a)$ according to
the system of characteristics (\ref{doc11}). Because of the large
number of short waves, the sum in eq. (\ref{doc22b}) can be taken as fast
oscillations average  with an adequate accuracy. Then the
expression for the superpotential gets the form
\begin{equation}
\label{doc22c} V=4\left[a^2-\frac{p}{8a^2}\left
(1-\frac{p}{2}\right)-\frac{1}{8}\sum_{k}C_k^2\right],
\end{equation}

The existence of an inner classically available region requires a
realization of the condition $V(a)<0$ in the  mentioned interval $0\le a
\le a_0$. (Let us remember --- see eq. (\ref{doc6}) --- that the  total
energy $E$ vanishes for the closed universe.) The potential barrier
should begin at $a=a_0$ , i. e. $V(a)>0$ at  $a\ge a_0$. Let us
notice that the vanishing  of the factor ordering parameter  ($p=0$) allows us
to get an
 inner classically available region only at the expense of the
sum $\sum C_k^2$. But in that case, a  potential barrier necessary
for a further development  is absent. So, we choose the factor
ordering $p=1$ in the following.

Owing to the amplitude of short waves begin to increase quickly
under the  barrier, we assume that their total energy is negligible in
the considered region $V(a)<0$, and the  evolution of the Universe is
determined  only by the first two terms in the r.h.s. of eq.  (\ref{doc22c}).
 In this case the
dynamics of the amplitudes of the  short waves  can be given
explicitly. The phase in eq. (\ref{doc20}) is

\begin{equation}
\label{doc22ed}
\Phi(a)=\frac{k}{2}\arcsin\left(\sqrt{1-\left(\frac{a}{a_0}
\right)^4}\right).
\end{equation}
 Here, $a_0$ can be obtained from the condition
\begin{equation}
\label{doc22d} V(a_0)= 4\left[ a_0^2-\frac{1}{16a_0^2}\right]
=0,\qquad a_0=\frac{1}{2}.
\end{equation}

 A case when  the total energy of the modes is not negligible in the
 expressions  for $V$ and $F$   can be similarly
 considered.  But then  the final expression becomes essentially
 complicated.

\section{ Modes of the scalar field under the barrier}

The superpotential $V$ is positive at $a>a_0$. This  means that the
evolution of the WF occurs under the barrier. At this range,
the problem is defined by the system of characteristics (\ref{doc13}),
and we have for $q_k$ the following equation
 \begin{equation}
\label{doc25} \frac{d^2q_k }{d a^2}= \frac{4k^2q_k}{F_e^2}+
\left(\frac{2}{a}-\frac{F_e'}{F_e}\right) \frac{dq_k }{d a}.
\end{equation}

This equation has an inverse sign at first term on the right in
comparison with eq. (\ref{doc19}) for the classically allowed
region. It changes the  solution in principle, but for a
large parameter $k^2$ we can find  a solution  by a  method
similar to the one used for eq. (\ref{doc19}) $$ q_k=C_ka\cosh\Phi_e(a);$$
\begin{equation}
\label{doc22e}
 {\rm A}_k=\frac{C_k}{2ak}\sinh\Phi_e(a);
\end{equation}
$$ \Phi_e(a)=2k\int\limits_{a_0}^a\frac{da}{F_e}.$$ Here the
conditions of continuity of $q_k$ and ${\rm A}_k$  at the point $a=a_0$
were used for the  selection of the  constants of integration. Both in
section 3 and here, the correctness of the solution (\ref{doc22e}) is
determined by the  requirement of $F_e'/F_e\ll k$. This  requirement is
checked by substituting  of the  obtained solution in $F_e$ of  eq.
(\ref{doc13}). There $$F_e=-\sqrt {\sum_{k}C_k^2+4a^2}.$$ In the
last formula the term containing the  factor ordering, that plays a part only
close to zero, is ignored.

Because of the  smallness of the sum $\sum_{k}C_k^2$ under the
barrier this term can be neglected in the  expression for $F_e$,
and we suppose with a necessary accuracy
\begin{equation}
\label{doc*} F_e(a)\simeq -2a.
\end{equation}
At that $a$ the  phase $\Phi_e$ has the form
$$\Phi_e=-k\ln\left(\frac{a}{a_0}\right),$$ and leaving only
growing modes in solutions at $a>a_0$ we obtain
\begin{equation}
\label{doc**}
q_k=\frac{C_ka}{2}\left(\frac{a}{a_0}\right)^k,\qquad{\rm
A}_k=-\frac{C_k}{4ka}\left(\frac{a}{a_0}\right)^k.
\end{equation}
\par\bigskip
In that solution the amplitude of short waves catastrophically
increases with growth of $k$ but not under the exponential law.
The latter  is realized in the case of a slow variation of $F_e$
in eq. (\ref{doc22e}). It is similar to the Rubakov-effect of
catastrophic  particle creation, see  ref. \cite{Rubakov}.

A final value of an amplification of the amplitude of short waves
is determined by a ``time" of   the WF stay under the barrier.
Let us notice that we consider a self-consistent problem of
amplification of short waves (an analogue of particle creation).
It means that the energy of the  increasing modes under the barrier
controls the  form of the barrier itself  (see Fig. 1).

\begin{figure}
\begin{picture}(50,160)
\put(-200,-280){\includegraphics{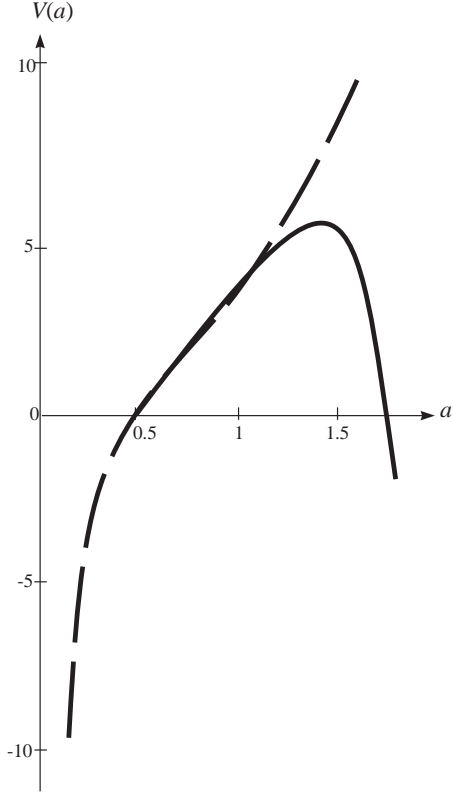}}
\end{picture}
\par\bigskip
\par\bigskip
\par\bigskip
\par\bigskip
\par\bigskip
\par\bigskip
\par\bigskip
\par\bigskip
\par\bigskip
\par\bigskip
\caption{\it{Form of the potential barrier without modes influence
(dashed line) and with taking them in account (solid line).}}
\end{figure}

Let us consider the equation for the action of the  WF under the barrier
(\ref{doc13})
\begin{equation}
\label{doc}
 \frac{d S_e }{d a}=\frac{V }{F_e};\qquad F_e\simeq -2a;\qquad
 V=4a^2-\frac{C_k^2}{4}\left(\frac{a}{a_0}\right)^{2k}.
\end{equation}
Only a ``potential" energy of increasing modes is  present in the
superpotential $V$, but here again it is not cancelled by their
``kinetic" energy as it was realized in (\ref{doc*}).

As it follows from $V(a_0, a_1)=0$, an external boundary of  the
tunneling region is determined by the equation
$$V=4a_1^2-\frac{C_k^2}{4}\left(\frac{a_1}{a_0}\right)^{2k}.$$

The final value of amplification of the amplitude of the modes is defined
by substitution of $a_1$ in eq. (\ref{doc**}).

Further, the Euclidean action $S_e(a_1)$ has a form

\begin{equation}
\label{1}
 S_e(a_1)=\frac{1 }{16}\sum_k
 C_k^2\left(\frac{a_1}{a_0}\right)^{2k}\left(\frac{1}{k}-1\right).
\end{equation}
Subject to $k\gg 1$ we have $S_e=-a_1^2$. Hence, a probability of
tunneling of the  WF through the barrier is given as

\begin{equation}
\label{11} w=\exp(-2|S_e|)=\exp(-2a_1^2)
\end{equation}
at $w\ll 1$.  The formula $$ \label{12}
w=\frac{\exp(-2a_1^2)}{1+\exp(-2a_1^2)}$$ is  more exact in the  case
 $w<1$.

\section{ Conclusions}

The self-consistent problem of the  evolution of the closed Friedmann
Universe with an amplification of wave perturbations of a scalar
field under a barrier is considered. Within the framework of a
semiclassical approximation of the Wheeler-DeWitt equation it is
shown that the increase of the amplitude of short waves under the
barrier controls the barrier shape, i.e. leads to both a reduction
of the barrier and fast output of WF from under the barrier. In
this sense the problem is  self-consistent. The process of tunneling
and the increase of the  amplitudes  of short waves is defined here
 by the assignment of their initial spectrum in
the inner classically allowed  region.  If it is required after
tunneling in the Universe to receive a spectrum close to the flat
one, the initial spectrum should be almost exponentially
suppressed in the  area  $k\rightarrow\infty$. It is qualitatively
equivalent to the  presence of an effective temperature in the  inner
classically allowed  region.

A more detailed discussion
of these problems obviously requires  a consideration of the  quantum
nature of the
 problem for short-wave perturbations of a scalar
field.

\medskip

{\bf ACKNOWLEDGEMENT}
\par
The authors thank A. A. Starobinsky  and M. V. Sazhin for
information about papers of V. A. Rubakov et al. and  H. Kleinert
for useful comments. H.-J.S. acknowledges financial support from
the HSP III-program. The work is done within the project  KR-154
 of the International Science and Technology Centre (ISTC).

\end{document}